\documentstyle[preprint,aps]{revtex}
\def\ra{\rightarrow}
\def\Ls{\Lambda_s (1405)}
\def\K{\bar{K}N}
\begin{document}
\preprint{\vbox{\baselineskip=14pt
   \rightline{UH-511-869-97} \break
   \rightline{July 1998} \break
    \rightline{revised, March 1999}
  }}

\draft
\title{$\Ls$ and Negative Parity Baryon States}

\author{S. Pakvasa\thanks{email address:  pakvasa@uhheph.phys.hawaii.edu}
and S.F. Tuan\thanks{email address:  tuan@uhheph.phys.hawaii.edu}}
\address{Department of Physics \\
 University of Hawaii at Manoa \\
              Honolulu, HI 96822 }


\maketitle
\begin{abstract}

          After a brief historical background on the $\Ls$,
          we revisit the 25 year old controversy on whether
          $\Ls$ is a $qqq$ (L=1) three quark state or a $\K$
          quasi virtual bound state.  This work is stimulated by the recent
          suggestion of Isgur that s be treated as a heavy quark in heavy
          quark effective theory HQET.
          We re-examine the empirical evidence for minimal
          mixing amongst singlets and octets in negative parity baryon
          states, with a possible dynamical origin in the opening
          of inelastic threshold channels. Finally, we suggest
          that $\Ls$ belongs to a class of hadrons which are
          described simultaneously  as $qqq$ or $q \overline{q}$ states
          and as hadronic bound states.

 \end{abstract}


\newpage   

  The $\Ls$ (=uds, similarly for $\Lambda_c$=udc, $\Lambda_b$=udb)
state with $I=0, J^P = 1/2^-$ was predicted in 1959\cite{Dalitz}
by extrapolating into the unphysical region below $K^-p$ threshold for $K^-p
\ra \pi+ \Sigma$ reaction amplitude, using zero effective range S-wave scattering 
lengths obtained from $\K$ data above threshold. It was discovered 
experimentally some two years later by the LBL Bubble Chamber Group \cite{Review}. As
attested by the compilation in for instance the latest Particle Data Group
book \cite{Review}, refined effective range analysis with accumulated $\K$ data have
always been able to reproduce width and mass parameters of $\Lambda_s$ rather
well. Furthermore there is the telltale signal of association with the $\K$
threshold in that the $\Ls$ resonant shape cut-off sharply at this 
threshold. The question of whether $\Ls$ is described correctly as a $\K$ bound state
or a three quark state has been discussed extensively with no definite conclusion \cite{Dalitz1}.
There is the suggestion from the chiral cloudy bag model \cite{Deloff}
that the $\K$ configuration dominates at the 90\% level over
the three quark configuration, and is supported by more recent work \cite{Kaiser}.
Hence it was never seriously doubted that $\K$ closed channel was 
significant in the formation of $\Ls$, and in fact it was a {\bf shock}
to have to think of this state in another way as discussed below. 

Nevertheless it has been recognized that quark model classification is
fundamental for {\bf all} hadronic matter, hence $\Ls$ must have assignment
in the quark model \cite{See}. The standard $(qqq, L=1^-)$ SU(6) 
70 representation of the quark model accommodates $\Ls$ and a companion 
$\Lambda_s(1520)$ in $J^P = 3/2^-$ as the (1,2) and (1,4) (SU(3), 2J+1) members of
\begin{equation}
   (70, L=1^-) = (8,6)^- + (10+8+8+1,4)^- + (10,2)^- + (8+8+1,2)^-.
\end{equation}

There has however been a number of problems for this uds classification of $\Lambda_s$
states, which can be stated as follows:-
\begin{description}
\item[(i)] The {\bf L.S} Puzzle. All states which have been assigned to multiplets
in (1) have very small spin-orbit {\bf L.S} contributions \cite{See,Feynman,Rujula,Isgur1}
except for the {\bf large} $\Lambda_s(1520)$ - $\Ls$ splitting of 115 MeV. 
Indeed a relativized quark
model proposed by Capstick and Isgur \cite{Capstick1} naturally ``explains" the 
{\bf smallness} of
spin-orbit interaction thus exacerbating the $\Lambda_s (1520)/\Ls$ non 
degeneracy anomaly. However, the same authors \cite{Capstick1} concede that $\Ls$ has 
unusually large couplings to $\K$ channel, with respect to which it is an
S-wave resonance. Coupling would naturally shift predicted uds state towards
or even below $\K$ threshold, which is where it is found. As argued by
Capstick \cite{Capstick2},
whenever a quark state is close to a threshold $[\bar{K}N]$
like $\Ls$, it is inconceivable that the {\bf narrow resonance approximation}
\cite{Isgur1,Capstick1} used in the calculation of masses will give correct answers - unless
by chance!
\item[(ii)] At its observed mass the $\Ls$ can decay only into $\Sigma \pi,
\Lambda\gamma$, and $\Sigma\gamma$; one way to test whether it can be interpreted
as a mass-shifted version of the uds quark state is to check, in addition to
the known $\Sigma\pi$ amplitude, the two radiative decay amplitudes against
those expected for the quark state interpretation. Unfortunately the earlier
evidence here for the quark interpretation \cite{Darewych} has been vitiated somewhat
by the 1991 Burkhardt isobar fit \cite{Burkhardt} which reported solutions for the 
branching ratios of $\Ls \ra \Lambda\gamma, \Sigma\gamma$ about an order of
magnitude smaller than those arrived at in \cite{Darewych}. There has
been some recent work
\cite{Moore} in this area, but no definite conclusions were reached on this issue.

\item[(iii)] To underline the gravity of the $\Ls$ problem, Jaffe  \cite{Jaffe}
has made the drastic suggestion that the large splitting with $\Lambda_s(1520)$ could be better
understood by re-assigning $\Ls$ to be a hybrid baryon (uds)g where $uds$
is in $1/2^+$ and g is the gluon. It seems to us, however \cite{Capstick2} that
a shift of less than 300 MeV for a hybrid excitation [the (uds) $1/2^+ \ \Lambda_s$ is at 
1116 MeV] is too small, and it is only the bag model's treatment of `constituent
gluons' which yields these low energies. None of the other low-lying hybrid
states predicted by the bag model have as yet surfaced in nature. If it costs only
300 MeV to excite the glue, we would have seen several hybrids in the non-strange 
spectrum, which is much better known. Although it has been speculated
that the Roper resonance N(1440) with $J^P = 1/2^+$ is a nonstrange hybrid 
baryon, one needs then to ask where is the even lower mass $qqq$ state with
those same quantum numbers? In the quark model \cite{Capstick1} this $qqq$ state 
with $I =1/2$ and $J^P = 1/2^+$ is predicted at about 1550 MeV. Here again the mass
prediction for this $3q$ state resorted to the use of the narrow resonance
approximation where in actuality we know that N(1440) has a large width to
$N\pi$ [60-70\% of a total width $\sim 350$ MeV]. Decay channel couplings probably
shifted the quark model prediction at 1550 MeV down to the Roper mass of
1440 MeV, a shift compatible with the 115 MeV shift downward of the
$\Ls$ from its quark model predicted value of 1520 MeV.  Jaffe
\cite{Jaffe} has recently suggested that the low mass (uds)g applies
only to unitary singlets where color magnetism is important.  However,
the need for a $J^P = 1/2^-$ uds state near $\Lambda_s (1520)$ remains to handle the
{\bf L.S} puzzle. 
\end{description}

     In a very stimulating recent contribution Isgur \cite{Isgur2} argues that recent data
from the $\Lambda_c$ system now strongly indicates that the $\Ls$ is in fact
a uds system, thus giving a new twist to the 25 year old $qqq (L=1)$ vs.
$\K$ controversy. 

In brief, his argument runs as follows:

\begin{description}
\item[(1)] A recent CLEO result \cite{CLEO} gives
\end{description}
\begin{eqnarray}
\frac{\Lambda_s(1520) - \Lambda_s(1405)}
       {\Lambda_c(2625) - \Lambda_c (2593)}  \ = 3.45 \pm 0.17 .
&  \quad  \quad  \quad \quad  \quad  \mbox{(2)}  \nonumber 
\end{eqnarray}
This is to be compared with the HQET prediction
of $m_c/m_s$.  The agreement can be considered to be reasonable when
constituent masses are used.
\begin{description}
  \item[2)] He then argues as follows:-
        \begin{enumerate}
            \item[a)] He inverts the argument to say that because (2) is true,
           ipso facto it is ``proof" that s-quark must be heavy in the HQET
           sense.
        \item[b)] In HQET, spin structure of $\Ls$ is fully determined.
        \item[c)] Therefore $\Ls$ is a 3q (uds) state and {\bf not} 
         a  $\K$ quasi bound state.
          \end{enumerate}
\end{description}

However the agreement in Eq. (2) is not quite as good as the other hyperfine
splittings \cite{we} e.g.
\begin{eqnarray*}
\frac{K^* (892) -K}{D^* (2010) - D} & =  2.830 \pm 0.015      \quad
\quad   \quad     \quad  \mbox{(3a)} \nonumber   \\
\frac{\Sigma^* (1385) - \Sigma}{\Sigma_c^* (2520) - \Sigma_c (2455)}  & = 
2.92 \pm 0.07 .      \quad    \quad      \quad \quad     \ \ \mbox{(3b)}
\end{eqnarray*}
The agreement in Eq. (2) is noticeably poorer than in Eq. (3) leading us
to conclude that the $\Lambda_s (1520) - \Lambda_s (1405)$ sector
remains qualitatively different and that our understanding of this
system is still incomplete. The errors in (2) and (3) are experimental
where we have used the PDG values.  

     The $\Ls$ as a difficult classification case for the $[70, 1^-]$ quark
baryon spectrum, has been an important stimulus in search for other deviations
from the standard quark model prediction \cite{Isgur1,Capstick1}. 
It has been found that there is  
increasing empirical support for an (8,2) $\eta$ octet of 
$1/2^-$ states\cite{Pakvasa,Tuan} associated with 
$\eta +N [N(1535)], \eta + \Lambda [\Lambda (1670)], 
\eta + \Sigma [\Sigma (1750)]$ S-wave threshold interaction. 
Although the $\eta$ + $\Xi$ member has yet
to be identified, such states within say 50 MeV of the appropriate thresholds
could satisfy an {\bf{unmixed}} Gell-Mann-Okubo octet mass formula to high accuracy.
The experimentally known members of the $\eta$-baryon octet all have 
significant coupling to the appropriate $\eta$+baryon channel, in the range 15 to
55\% in decay partial width. As in the case of $\Ls \leftrightarrow \K$ 
\cite{Capstick2} these states are close to the 
$\eta$ baryon thresholds and have significant couplings to the appropriate $\eta$-baryon
channels, and there could be sizeable mass shifts 
from the predictions of the $[70,1^-] \ qqq$ model
\cite{Isgur1,Capstick1}. Indeed the S-wave  $1/2^- \ \eta$ octet 
violates a rule suggested by Feynman \cite{Feynman} for
(mass)$^2$ to wit:-
\begin{eqnarray*}
          \mbox{Sign}  [\Sigma^2 - \Lambda^2] = \mbox{Parity of the state.} \quad  \quad  
     (4) \nonumber
\end{eqnarray*}
whereas the Isgur-Karl model\cite{Isgur1} largely satisfies this rule. 
(Established $J^P= 3/2^-,5/2^-$ octets \cite{See,Tuan} also appear to satisfy this rule.) 
It is now worthwhile to re-examine the classification of negative parity baryon
states of the standard quark model \cite{Capstick1} in the light that nature has provided
us with unmixed (in the octet mass formula sense) $\eta$ octet of $J^P = 1/2^-$
states as well as the earlier $J^P = 3/2^- \ \gamma$ octet \cite{Martin} which has now 
received impressive experimental support \cite{Tuan}. The methods used 
\cite{Deloff,Darewych}
to determine the composition of $\Ls$ as a superposition of three-quark
and $\K$ configurations should now be applied to the $\eta$-baryon $1/2^-$ octet
to determine their compositions as superpositions of three-quark 
and $\eta$-baryon S-wave ``molecular" configurations. Bugg \cite{Bugg} has stressed that
in connection with the $K\bar{K}$ and $\eta\eta$ thresholds in the $0^+$ sector inelastic 
thresholds can influence both the shape of a resonance and move its mass
around by a substantial amount. We suggest that the same is true in the 
baryon sector not only for $\Ls$, but also for the 
$\eta$-baryon octet and indeed the $\gamma$ octet where the $J^P= 3/2^-$ member
 N(1512) could have
a dynamical origin due to the opening of the inelastic $\rho$-N S-wave threshold
as proposed many years ago by Ball and Frazer \cite{Ball}.  It has been
suggested that the mechanism for $\Lambda_s (1405) - \Lambda_s (1520)$
mass difference is driven largely by N-$\Delta$ and K-K$^*$ mass
differences through mixing with nearby threshold.  This is an
interesting idea, but beyond the scope of our Letter here.

To summarize, we see that there is some evidence that treating
the s-quark as heavy in HQET \cite{Isgur2}, together with the
corrobative evidence in the meson spectra \cite{Suzuki}, points towards
the interpretation that $\Lambda_s (1405)$ is 3q (uds) in nature.
[Although the reason why it is permitted to treat s-quark as heavy, when
$m_s$ is not large compared to either $\Lambda_{QCD}$ or $m_u$, $m_d$
(in the constituent basis), remains mysterious.]  On the other hand,
perhaps even stronger evidence \cite{Review,Deloff,Kaiser} exists that
it is a $\overline{K}N$ quasi virtual bound state in the molecular
sense.  Hence it seems extremely unlikely that either picture can be
``wrong''.  

It seems to us that whereas the description in terms of 3 quarks is
clear cut for many baryons; at the other extreme there are baryonic
states which are quite clearly not well described as quark states \cite{See1}.
For example, the deuteron is
a clear example of a bound state of neutron and proton as demonstrated
by powerful arguments given by Weinberg \cite{weinberg}.  Furthermore,
there is no evidence for other members of the six-quark multiplet.  We
submit that 
a state such as $\Lambda_s (1405)$ lies somewhere in the middle.  Namely,
whereas it is a three quark state and is classified as such; many of its
properties, such as production (and formation), mass shift, coupling to
various channels are heavily influenced by the proximity of the nearby 
$\overline{K} N$ threshold and the S-wave nature of the coupling.

A number of other hadronic states have similar behaviour and are
influenced strongly by nearby thresholds.  These are:
\begin{enumerate}
\item[(I)]
As already mentioned, the formation of the $\eta$ and $\gamma$ octets,
is also strongly influenced by nearby inelastic S-wave channels.  They
are furthermore characterized by mixing purity as a distinguishing feature.

\item[(II)]
The axial-vector kaons seem to be mixed in ways which favor specific
decays, e.g. $K_1 (1270) \ra K \rho$ or
$K_1 (1400)\ra K^*\pi$ in S-waves as noted long ago\cite{colglazier}.
Similar mixing was also found in $D^{**}$ states as noticed  by de 
${\rm R\acute{u}jula}$ et al. \cite{Rujula1}; and a similar 
tendency in baryons was observed in the
Isgur-Karl paper \cite{Isgur1}.

\item[(III)]
The $a_0 (980)$ and $f_0 (980)$ appear to be prominent effects on the
I=1 and I=0 $K \overline{K}$  systems near threshold.  Their assignment as
$q \bar{q}$
states is somewhat problematic because of their low $\gamma \gamma$
widths \cite{barnes}. There are recent proposals for a scalar nonet
 for both $q \bar{q}$  \cite{lucio} as well as non-molecular
 $q q \bar{q} \bar{q}$ \cite{achasov}; but the $K \overline{K}$
 seems to be the dominant component\cite{weinstein}.

\item[(IV)]
The $E/\iota (1420)$ (an I=0 $K \overline{K} \pi$ resonance) has a mass such that
$a_0 (980)$ and the two $K^* (890)$ bands in the Dalitz plot all cross at
the same point.  Perhaps it is a ``molecule'' of its three final-state
particles, in which each pair resonates simultaneously.

\item[(V)]
There is also the $\xi (2220)$ \cite{bai} which may be an example of a
state with simultaneous presence of $gg, q \bar{q}, qq \bar{q} \bar{q}$, and 
$q \bar{q} g$ components.  Given its proximity to the $\Lambda
\bar{\Lambda}$ threshold, this could be the key channel.
\end{enumerate}

Recently, Isgur \cite{nathan} has developed a formalism on the impact of
thresholds on the hadronic spectrum beyond the adiabatic approximation.
We feel that this formalism  could and should be applied to a number of
phenomenological issues raised here.  In particular the threshold shift for
$\Lambda_s (1520) - \Lambda_s (1405)$ problem discussed by us
qualitatively, can now be given a quantitative basis. 
Hence this formalism has a number of other
potential applications, as listed above under (I) to (V), where it makes
sense to view hadronic states simultaneously as states containing the
lowest number of quarks (e.g. $q \bar{q}$ or $qqq)$ and as bound states or
resonances involving specific channels (and hence involve additional 
$q \bar{q}$ pairs).

One of us (S.F.T.) wishes to thank Nathan Isgur and Simon Capstick for
useful communications and we thank Rob Kutschke, Mahiko Suzuki, but
especially Jon Rosner for  useful suggestions and discussions.  This
work was supported in part by the US Department of Energy under Grant
DE-FG-03-94ER40833 at the University of Hawaii at Manoa.

\end{document}